# NELIOTA LUNAR IMPACT FLASH DETECTION AND EVENT VALIDATION

Alexios Liakos[(1)], Alceste Bonanos[(1)], Emmanouil Xilouris[(1)], Ioannis Bellas-Velidis[(1)], Panayotis Boumis[(1)], Vassilis Charmandaris[(2)], Anastasios Dapergolas[(1)], Anastasios Fytsilis[(1)], Athanassios Maroussis[(1)], Detlef Koschny[(3), (4)], Richard Moissl[(3)], and Vicente Navarro[(5)]

[(1)] *Institute for Astronomy, Astrophysics, Space Applications and Remote Sensing, National Observatory of Athens, Metaxa & Vas. Pavlou St., GR-15236, Penteli, Athens, Greece, Email: alliakos@noa.gr*
[(2)] *Department of Physics, University of Crete, 71003 Heraklion, Greece, Email: vassilis@physics.uoc.gr*
[(3)] *Scientific Support Office, Directorate of Science, European Space Research and Technology Centre (ESA/ESTEC), 2201 AZ Noordwijk, The Netherlands, Email: Richard.Moissl@sciops.esa.int*
[(4)] *Chair of Astronautics, Technical University of Munich, 85748 Garching, Germany, Email: Detlef.Koschny@esa.int*
[(5)] *European Space Astronomy Centre (ESA/ESAC), Camino bajo del Castillo, s/n, Urbanizacion Villafranca del Castillo, Villanueva de la Cañada, 28692 Madrid, Spain, Email: Vicente.Navarro@esa.int*

## ABSTRACT

NELIOTA (NEO Lunar Impacts and Optical TrAnsients) is an ESA-funded lunar monitoring project, which aims to determine the size-frequency distribution of small Near-Earth Objects (NEOs) via detection of impact flashes on the surface of the Moon. A prime focus, high-speed, twin-camera system providing simultaneous observations in two photometric bands at a rate of 30 frames-per-second on the 1.2 m Kryoneri telescope of the National Observatory of Athens was commissioned for this purpose. A dedicated software processes the images and automatically detects candidate lunar impact flashes, which are then validated by an expert user. The four year observing campaign began in February 2017 and has so far detected more than 40 lunar impact events. The software routinely detects satellites, which typically appear as streaks or dots crossing the lunar disk. To avoid confusing these events with real flashes, we check different available catalogs with spacecraft orbital information and exclude spacecraft identifications.

## 1 INTRODUCTION

Near-Earth Objects (NEOs) are defined as asteroids or comets that can cross the orbit of the Earth and potentially can cause serious damage on the surface of the planet (e.g. destroy infrastructure). Meteoroids are tiny objects on the order of up to one meter that are asteroidal or cometary debris. The majority consist mostly of stone (chondrites and achondrites), but many of them are stone-iron composites or consist only of iron. Known debris from asteroids and comets, when entering the atmosphere of the Earth, can be observed as meteor showers, while the randomly occurring meteors are characterized as sporadics.

Observations of small NEOs and meteoroids entering the Earth's atmosphere have certain difficulties such as the very limited covering area (order of a few thousands of km$^2$) and the fact that very small meteoroids do not generate any visible light makes them impossible to be detected. On the other hand, the Moon presents certain advantages for indirectly detecting small size NEOs and meteoroids by their impact flashes. In the absence of an atmosphere, the projectile directly impacts the lunar surface and a portion of its kinetic energy is converted to luminous energy. That transition produces light emission in a form of a flash that can be easily recorded from ground-based telescopes. Moreover, the maximum potential effective area for observations of flashes is the lunar side facing Earth, which is approximately 19 million km$^2$. However, the sunlit part of the Moon affects the observations (i.e. contrast increase) and, therefore, sets observational constrains regarding the phase of Moon during which the observations can be obtained. Hence, the total available nights for lunar observations per month strictly depend on the lunar phase. The greatest challenges that need to be overcome, when trying to detect impact flashes on the Moon, are the seeing fluctuation due to the Earth's atmosphere and their short duration. For the latter, sensitive and fast read-out cameras are needed in order to record lunar images at high speed frame rates.

Several regular lunar impact flash monitoring programs have been performed, for example by the MIDAS team in Spain [1], [4], [5], [6], by NASA's meteoroid office [3], by a team in New Mexico [7] and one in Morocco [8]. Other teams have performed shorter period activities, e.g. as summarized in [2]. The common characteristics of these campaigns are: a) observations using multiple small-sized telescopes (diameter of 30-50 cm) and b) unfiltered [1], [2], [4]-[8] or single band observations [3].

The "NEO Lunar Impacts and Optical TrAnsients" (NELIOTA) project began in early 2015 at the National Observatory of Athens (NOA) and is funded by the

---

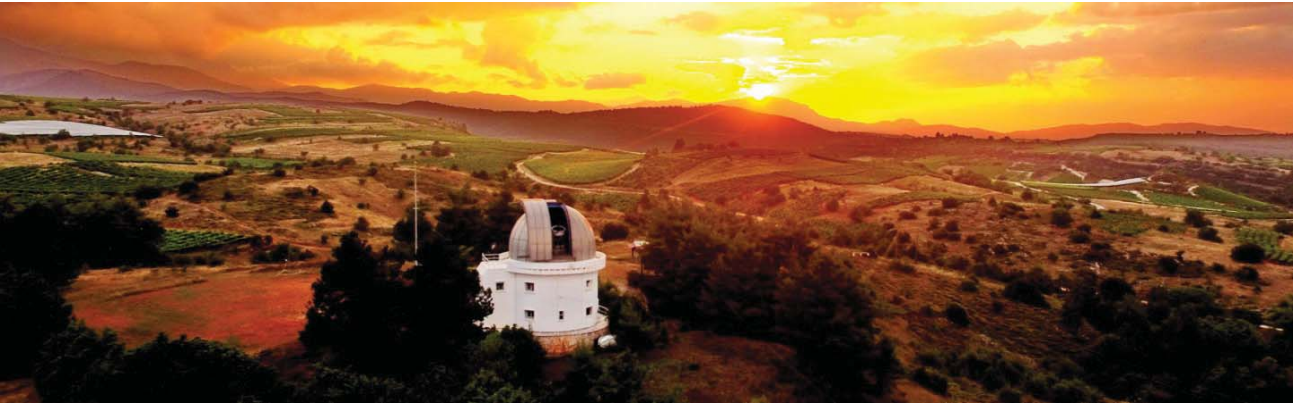

*Figure 1. Aerial view of the Kryoneri Observatory.*

European Space Agency (ESA). Its short-term goal is the detection of lunar impact flashes and the estimation of the physical parameters of the projectiles (mass, size) as well as those of the impacts (e.g. temperature, effects on the surface). The mid-term goal concerns the statistics of the frequency and the sizes of these objects to be used by space industry as essential information for the shielding of the space vehicles. For the purposes of the project, a dedicated instrumentation setup has been installed at the 1.2 m Kryoneri telescope[1] allowing observations at a high recording frame rate simultaneously in two different wavelength bands. In particular, the last capability is a specialty for a regular monitoring program with a telescope of this size, since it provides: a) the opportunity to estimate directly the temperature of the flashes [9], and b) to validate events using a single telescope. Details about the instrumentation setup and its efficiency/performance on lunar impacts can be found in [10]. The retrofit of the Kryoneri telescope was completed in June 2016, while the evaluation of the new equipment and the testing period of the project lasted until February 2017. The observing campaign started in March 2017 and is scheduled to continue until January 2021.

A dedicated software pipeline has been developed for the purposes of NELIOTA. The software performs the creation of the observation plan, the reduction of the lunar images, the subtraction of the lunar background, and the detection of the events. Subsequently, the events are inspected by an expert user and based on a validation flow chart are finally characterized as true, suspected or false events. For the latter, a cross check with satellite orbits is performed in order to reject the possibility for a slow moving and rotating satellite crossing the lunar disk to be characterised as a lunar flash.

## 2 INSTUMENTATION AND SOFTWARE

The observations are being carried out at the Kryoneri

[1] http://kryoneri.astro.noa.gr/

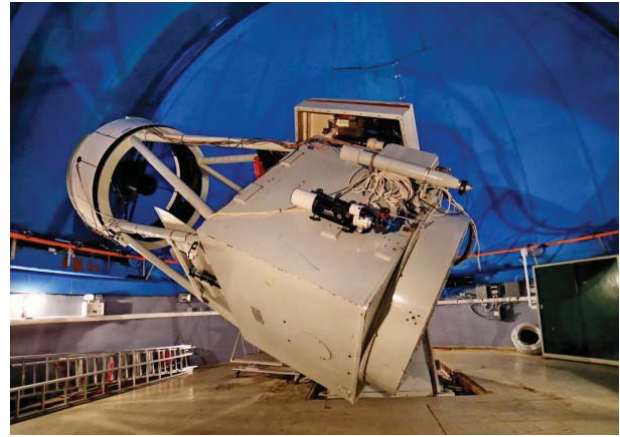

*Figure 2. The 1.2 m Kryoneri telescope.*

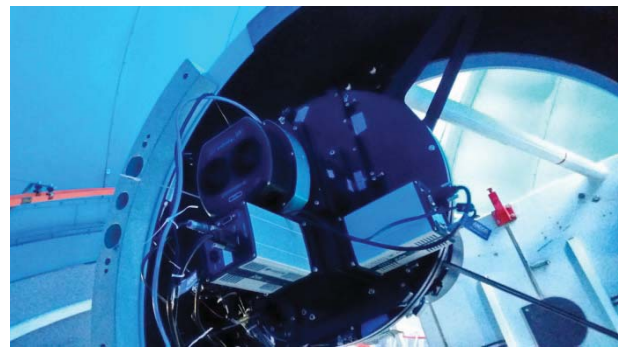

*Figure 3. The Prime Focus instrument of the Kryoneri telescope. The NELIOTA twin camera system consists of two identical sCMOS (grey devices) cameras installed on a beam splitter perpendicular to each other. Another CCD camera (black device) has been also installed for other purposes.*

Observatory (Fig. 1), which is located at Mt. Kyllini, Corinthia, Greece at an altitude of 930 m. The primary mirror of the telescope (Fig. 2) has a diameter of 1.2 m and a focal ratio of 2.8. Two twin front-illuminated sCMOS cameras (Andor Zyla 5.5) with a resolution of 2560×2160 pixels and a pixel size of approximately 6.5 μm are separated by a dichroic beam-splitter (cut-off at

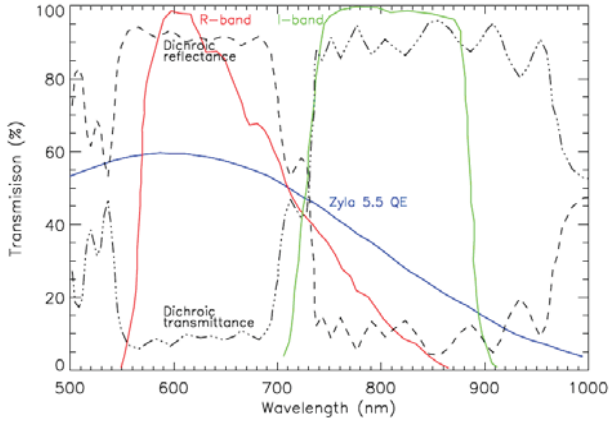

Figure 4. Filters (red solid line for the R and green solid line for the I photometric bands) and dichroic beam splitter transmission curves (dashed and dashed-dotted lines) combined with the QE response of the sCMOS detectors (blue solid line). The dichroic is centred at 730 nm with a throughput greater than 90%. Image was taken from [10].

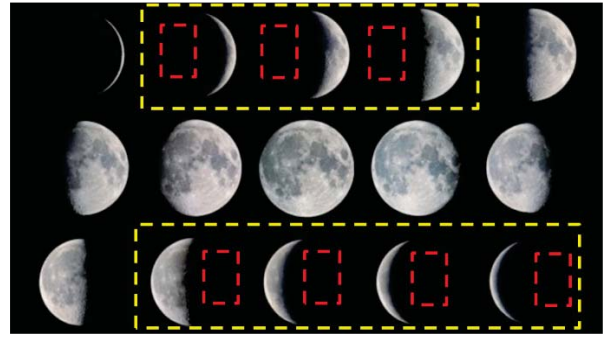

Figure 6. A full lunar period. Yellow rectangles denote the phases of the Moon when NELIOTA operates and the red rectangles the target lunar area.

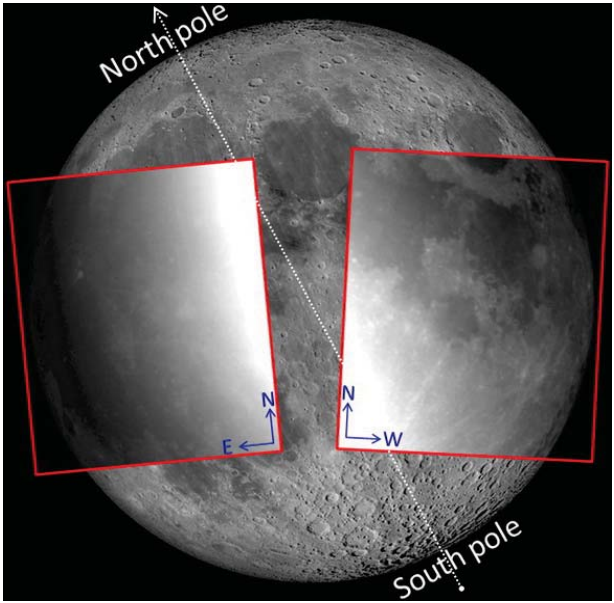

Figure 5. Examples of the field-of-view of the NELIOTA setup during the first (left) and the third (right) quarters of the Moon.

730 nm) and are mounted at the prime focus of the telescope (Figs 3, 4). Each camera is equipped with one filter of Johnson-Cousins specifications. In particular, the first camera records in the red ($R_c$) and the other in the near-infrared ($I_c$) pass bands, with the transmittance peaks being at $\lambda_R$=641 nm and $\lambda_I$=798 nm, respectively (Fig. 4). The field of view (FoV) of this setup is approximately 17.0×14.4 arcmin$^2$ (Fig. 5). Using the above set up, the typical observable lunar surface is approximately 3.5 million km$^2$ (at the average Moon-Earth distance) but it varies according to the Earth-Moon distance and the lunar phase. The cameras are connected via fibers to a server with a high-capacity storage array of 38 TB. Moreover, a Global Positioning System receiver to obtain the precise time (accuracy <0.2 ms) and a weather station have been installed for the needs of the NELIOTA programme.

A software pipeline has been developed for the purposes of the project. The software splits into four parts: a) Observations (NELIOTA-OBS), b) Reduction of data and detection of events (NELIOTA-DET), c) Archiving (NELIOTA-ARC), and d) Information (NELIOTA-WEB). The first three parts have been installed on the server located at the Kryoneri Observatory, while the fourth part has been installed on another server at NOA/IAASARS headquarters in Athens. More details regarding the observing equipment as well as for the whole infrastructure can be found in [10].

## 3   LUNAR OBSERVATIONS

The observations occur between the 0.10 and approximately lunar phases of 0.45 (i.e. from waxing crescent to the first quarter phases and from the last quarter to the waning crescent phases; 5-8 nights per month) at the non-sunlit (nightside) part of the Moon (Fig. 6). The upper limit of the lunar phase (see also Section 4) during which observations can be obtained depends strongly on the intensity of the glare coming from the sunlit part of the Moon. The upper limit of the lunar phase has been set at approximately 0.44 during the apogee and 0.46 during perigee. The effect of glare on lunar images is shown in Fig. 7.

The orientation of the cameras (Fig. 5) has been set in such a way so that the longer axis is almost parallel to the terminator during the whole year, i.e. it corresponds to the celestial north-south direction. Therefore, only the eastern-western hemispheres of the Moon can be observed but not its poles. The total covered lunar area by the NELIOTA setup depends on: a) the libration of the Moon, b) the inclination of the Moon axis with respect to the orbital plane, and (c) the varying distance of the Moon. The latter factor (c) is the most critical one since it concerns the size of the lunar area covered by the angular FoV of NELIOTA (i.e. the longer the Earth-Moon distance the greater the covered area). Therefore, every lunar observation covers a unique area. In total, all these factors allow to observe areas with latitudes up to ±50° from the equator and ±90° from the meridian.

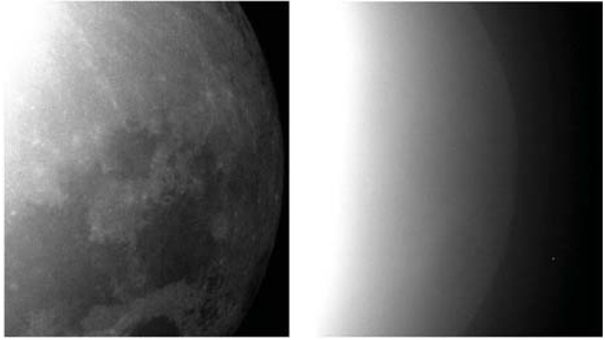

*Figure 7. Effect of glare on the non sunlit part of the Moon. The left image was taken during the lunar phase of 0.14 while the right image during the lunar phase of 0.44. On both images the left edge is not saturated. Contrast decrease is obvious for the right image and results in high background values that prevents the detection of faint flashes.*

The cameras are recording simultaneously at a frame rate of 30 frames-per-second (fps) in 2×2 binning mode. The selection of this binning mode was made in order to: a) increase the signal-to-noise (S/N) of the events, b) reduce the read-out time of the cameras (factor of 4), and c) decrease the data file size (factor of 4). Moreover, the choice of the binning was also based on the typical seeing of the observing site (approximately 1.5´´), so that the pixel scale (0.8´´ $pixel^{-1}$) is always smaller. The frame rate selection was based on the typical duration of lunar flashes (from 10 ms to several seconds) aiming mostly to detect the shortest ones. The exposure time of each frame is 23 ms followed by a 10 ms read-out time with the synchronization of the cameras to be better than 6 ms. Given the large telescope aperture, it is feasible to detect faint flashes (about 2 mag fainter than observed in previous campaigns; [1]-[8]).

Initially, the NELIOTA-OBS software validates the communication between all the components of the system (telescope, cameras, positioning system, weather station, and storage array). Then, its "planner" utility creates an observation plan for the night and splits the total available observing time into many lunar data "chunks" that last 15 min. The total available observing time is defined as the time during which: a) The Moon is above an altitude of 20° (limit set by the dome), b) the lunar phase is between 0.1 and 0.46, and c) the Sun is below the local horizon. The observations begin/end approximately 20 min after/before the sunset/sunrise.

The minimum duration of the observations is approximately 20 min (at low brightness lunar phases), while the maximum is approximately 4.5 hr (at lunar phases near 0.45).

At the end of each data chunk, a standard star is observed for the magnitude calibration of the flashes. Five images with each camera are typically obtained. Sky flat-field frames are taken before or after the lunar observations, while the dark frames are obtained directly after the end of them.

All raw data are stored in the storage array and they are kept until the disks are full. It should to be noted that due to the short exposure requirements approximately 40% of the total available observing time is devoted to the read-out time of the cameras (30%) and the repositioning of the telescope for the standard stars observations (10%).

## 4 DETECTION OF EVENTS

The NELIOTA-DET performs data reduction automatically using the median images of the respective calibration files (sky flat-field and dark images). These master images are used to calibrate the data of the Moon as well as those of the standard stars. Then, the software creates a mask for each data chunk, in order to (a) avoid searching for events beyond the lunar surface, and (b) to calculate the background to be subtracted from the original Moon images. For (a), the limb of the Moon has to be approximately 20 pixels from the edge. Then the software can determine a circular mask, which will be applied to all images to exclude areas outside the lunar disk. For (b), the software produces a time-weighted average image of the preceding images, i.e. the closer in time, the more weight is given to the image. This background will be subtracted from the current image before running the detection algorithm (more details can be found in [10], Section 5.2). It should be noted that the residual images still have a background that is connected with the lunar phase [10]. In particular, the standard deviation of the background of the residual image is larger when observing the Moon at higher brightness phases, while it becomes smaller for observations during lower brightness phases. At brighter phases, the standard deviation of the background becomes such that the software is not able to distinguish whether a detection comes from an external source (e.g. impact flash, cosmic ray) or from the fluctuation of the background itself. This results in multiple detections in almost every single frame. An upper limit of 7,000 ADUs (i.e. photon counts) per pixel of the lunar background during the observations has been roughly set for the subsequent nominal operation of the NELIOTA-DET. This limit corresponds to a lunar phase of approximately 0.45 with the exact value depending also on the Earth-Moon distance.

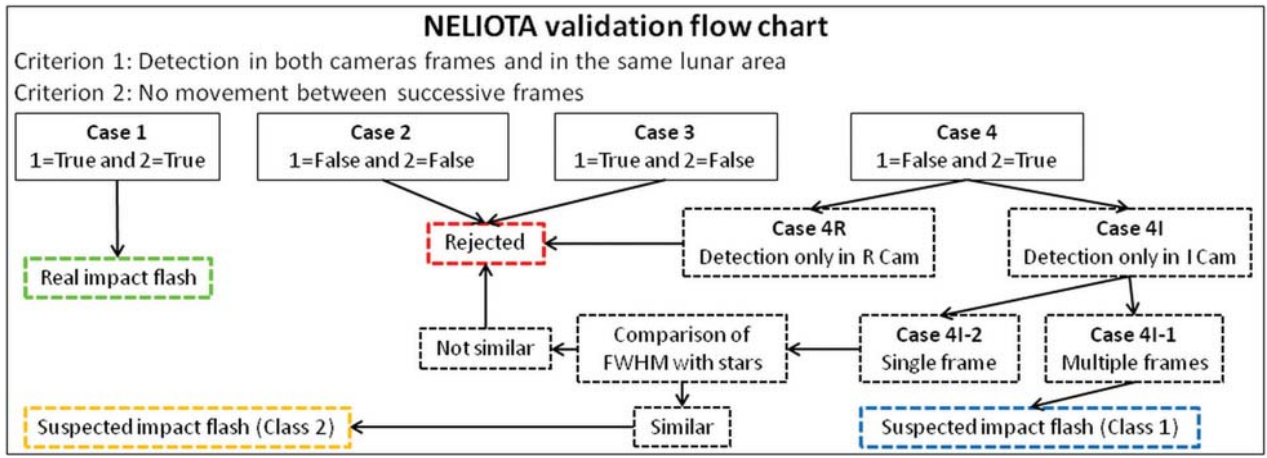

*Figure 8. Validation flow chart for the characterization of the events detected by NELIOTA-DET as implemented by the expert user. The criteria are noted on the upper part and all the possible characterizations are given in colored-edge text boxes.*

After the lunar background subtraction, the software searches for regions of more than 10 adjacent pixels, whose brightness is more than one standard deviation above the remaining background image. Once a potential event is detected, the software creates a directory that contains from both cameras regardless on which one the detection has been made: a) the frame(s) of the event along with the seven previous and the seven following ones, b) the residual frames and c) the background image used. It should be noted that when the NELIOTA-DET detects a potential event on one frame or on consecutive frames, this/these frame(s) is/are excluded from those used for the creation of the mean background image.

The reason for including seven frames before and after the detection of an event as well as the respective frames from the other camera is the detection threshold of the NELIOTA-DET. In particular, an event may be too faint to be detected by the software on a given frame but it may be brighter on the next one, on which the software indeed detects it. Therefore, for safety reasons, the previous and the subsequent frames have to be visually inspected by the expert user.

Moreover, it is very common that events are detected in the frames of one camera, but not in the corresponding ones of the other. Given that the impact flashes have relatively cool temperatures [9], it has been noticed that many of them are below the detection threshold of the software in the *R*-band frames. Therefore, for every event detection in the *I*-band camera, the corresponding *R*-band frames have to be again visually examined by the expert user.

## 5 VALIDATION OF EVENTS

Before proceeding to the validation procedure, the frequently used term "event" has to be defined. Therefore, an "event" is defined as whatever the NELIOTA-DET managed to detect. A detected event could be a cosmic ray hit, a satellite crossing the lunar disk, a field star very close to the lunar limb, and, obviously, a true impact flash. As mentioned before, all the frames that include events are stored in different directories along with the previous and successive seven ones from both cameras. Moreover, the software creates a list that includes all the detections in chronological order and for each of them performs a first characterization according to its shape (i.e. the software measures its roundness and radius) and its distance from the lunar limb. Subsequently, the NELIOTA expert user has to examine the events one by one and to validate them.

First of all, we set the validation criteria for characterizing the events. The first criterion concerns the detection of an event in the frames of both cameras and on exactly the same lunar area (i.e. same area of pixels). Therefore, it is feasible to search for an event at a specific area in the frames of one camera if it has been detected in the frames of the other. The second criterion is the lack of motion of the event between successive frames. There are four possible cases for the (non) satisfaction of these criteria, which they are addressed in detail below, while a schematic flowchart is given in Fig. 8.

*Case 1*: Both of them are fulfilled. The event is validated as a "Real lunar impact flash".

*Case 2*: None of them is fulfilled. The event is detected in multiple frames of only one camera (Fig. 9-upper panel). This case happens mostly when satellites are detected in the frames of only one of the cameras or when stars are very close to the lunar limb (Fig. 9-lower panel) are faint in one of the two pass bands. The event is characterized as false.

*Case 3*: The first criterion is fulfilled but the second is not. This case is frequently met when moving objects

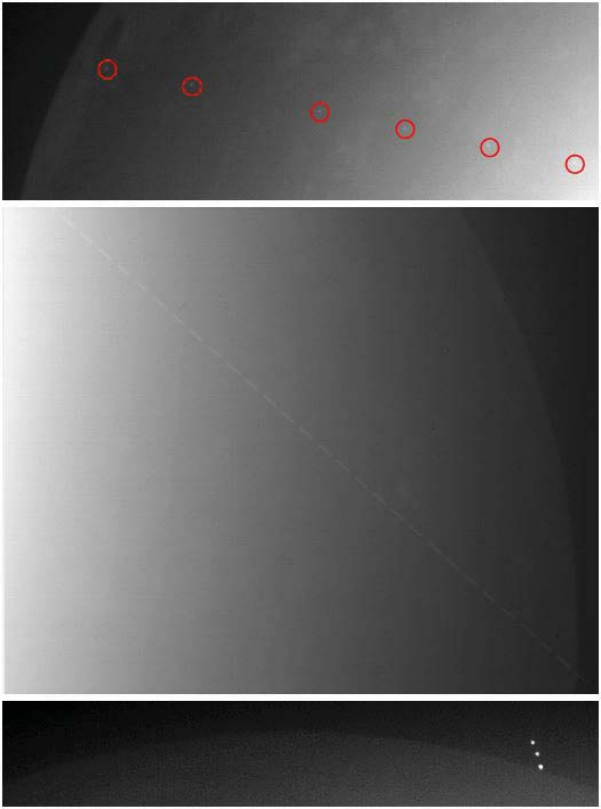

*Figure 9. Examples of moving artefacts detected by the NELIOTA-DET. The upper figure is an overlay of seven images of I camera with a time difference of 2 sec and shows a slow fly-by of a satellite in front of the lunar disk. The detection in the respective R frames was below the threshold. The middle figure consists of 29 successive images and shows a fast passage of a satellite in front of the lunar disk. This object was recorded in both cameras' frames. The lower figure is a superposition of three images with a time difference of approximately 60 sec between them and shows a starset over the lunar limb.*

(satellites, airplanes, birds) cross the lunar disk and are detected in the frames of both cameras (Fig. 9-middle panel). The event is characterized as false.

*Case 4*: The first criterion is not fulfilled, i.e. the event is detected in only one of the two cameras' frames. The latter produces two sub cases, namely Case-4R for the detection in the frames of *R* camera only and Case-4I for the detection in *I* camera only. This case is the most difficult and is related to the cosmic rays hits. In general, cosmic rays hit the sensors at random angles, producing, in the most cases, elongated shapes, which can be easily discarded (Fig. 10c). However, some of them hit the sensors almost perpendicularly providing round Point-Spread-Functions (PSF) like those of the stars (Fig. 10a, b). These two sub cases are addressed below.

*Case 4R*: According to the first results about the temperature of the lunar impacts flashes [9], the apparent magnitude of a flash in the *I* band is always brighter than in *R*, i.e. $R-I>0$. In addition, due to the Rayleigh scattering (i.e. the *R* beam scatters more than the *I* beam), events in the *I* filter camera can be detected more easily, in a sense that they exceed the lowest software threshold. This information provides the means to directly discard events that have been detected only in the frames of the *R* camera and not in the respective ones of the *I* camera. Therefore, the events of Case-4R are characterized as false.

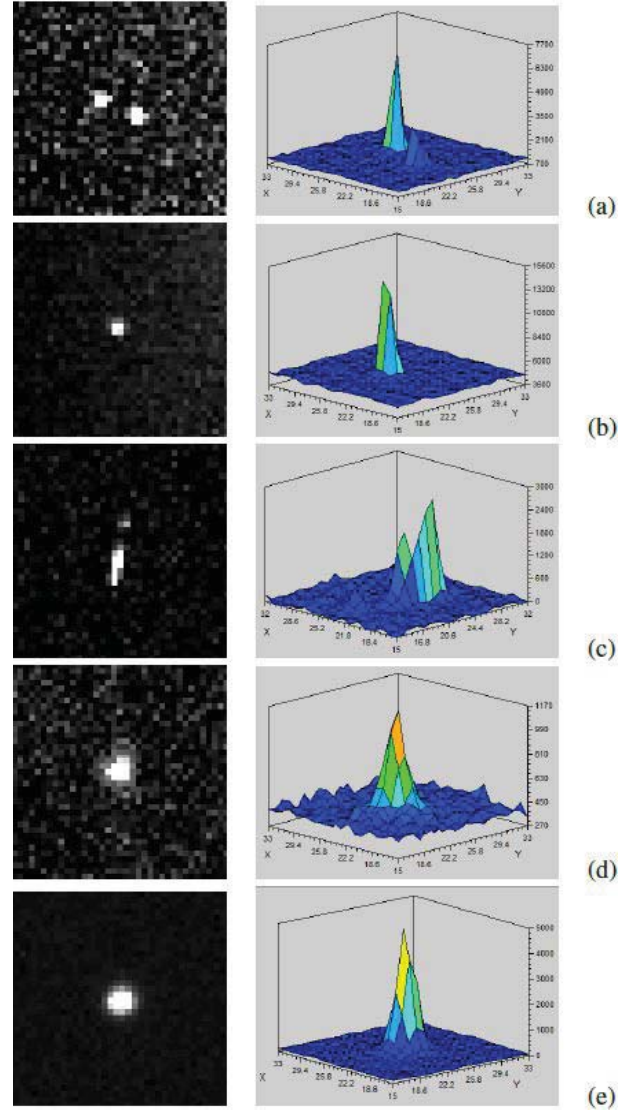

*Figure 10. Photometric profile examples of events. Left images show the events (40×40 pixels area), while the right images their respective profiles. The (a) and (b) images represent cosmic rays that hit the sensor almost perpendicularly producing round PSFs (i.e. star-like shapes). The (c) image shows another cosmic ray hit, but with another impact angle that produced an elongated profile. The image (d) illustrates a real impact flash, while image (e) the standard star used for the photometric calibration. The similarity in terms of FWHM is obvious for (d) and (e) (FWHM difference ~0.2 pixels), while a large difference (~1.5 pixels) is seen between (b) and (e).*

*Case 4I*: Taking into account the second criterion regarding the non-movement of the event between successive frames, the Case-4I is further split into two sub cases, which are described below.

*Case 4I-1*: The first sub case concerns the satisfaction of the second criterion, which means that the event is detected in exactly the same pixels of multiple successive frames and exhibits a round PSF. Cosmic ray hits last much less than the integration time of the image, therefore they are impossible to be detected in more than one frames. Therefore, Case 4I-1 events (i.e. multi-frame events detected in *I* camera only) are considered as "Suspected lunar impact flashes-Class 1". Class 1 denotes that the events have high confidence to be considered as true.

Case 4I-2: For this sub case, the detection has been made only in one frame of the *I* camera. This case is the most challenging one and further verification is needed. The reason for this is that cosmic rays, with intensities well inside the dynamical range of the sCMOS, hit the sensor almost perpendicularly producing round PSFs and, therefore, cannot be easily distinguished from the fast real impact flashes. As diagnostic tool the shape (i.e. the Full Width at Half Maximum - FWHM) of a star is used and compared with that of the event. In most cases, there are no field stars in the frame where the event is detected, so, that information comes from the standard stars observed between the lunar data chunks. However, on one hand, the standard stars have similar airmass values with that of the Moon, but, on the other hand, the Moon is observed typically at high airmass values where the seeing fluctuations are quite strong [10]. So, the event, if it is real impact flash, is not expected to present exactly the same FWHM value as that of the star, but to vary in a certain range. For this, the FWHM values of 42 real impact flashes and those of the standard stars used for their magnitude calibration were correlated. For each case the differences in terms of FWHM (pixels) were derived. This correlation is shown in Fig. 11 and produces a mean of 0.56 pixels and a standard deviation of 0.71 pixels. An example of this comparison can be seen in Fig. 10b, d. Therefore, according to these results, the round PSF single I-frame events that have difference ±1.3 pixels in FWHM from that of the standard stars are considered as "Suspected lunar impact flashes-Class 2", with Class 2 to denote that the events have low confidence to be considered as real.

Another difficult case that has to be checked very carefully concerns the extremely slow moving satellites (i.e. geostationary), which produce round PSFs very similar to the true flashes. In cases of events similar to that shown in Fig. 9-upper part, the satellite moves and spins around extremely slowly, so it can be detected in only one frame every a few seconds. Particularly, when the spinning period of the satellite is significantly greater than the integration time of the cameras (23 ms), then it cannot be detected in more than one successive frames. However, when the satellite crosses a significant portion of the lunar disk, the detections present a clear time pattern (e.g. one detection every two seconds), so it is feasible to reject those events. The latter, however, depends strongly on the path of the satellite in front of the lunar disk. According to the angle between the Sun-solar panels surface-observer the satellite can either be detected in one or in both cameras' frames. Depending on the altitude and the speed of the slow moving satellite, its crossing in front of the lunar area, that is covered by the FoV of the NELIOTA setup (Fig. 5), may last from several seconds up to a few minutes.

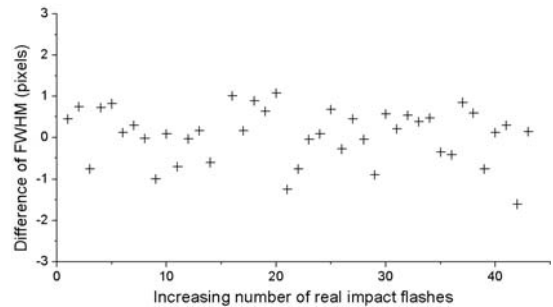

*Figure 11. Differences between the FWHM of standard stars and real impact flashes in I-band filter for 42 validated cases (i.e. detected in both cameras' frames) This correlation is used for the characterization of single frame I-band events. These flashes have been checked for possible satellite misidentification.*

Satellites crossing the FoV will trigger the detection software. Most satellites will be easily recognized since they will move during the exposure time and leave a trail. Normally they also do not change their brightness quickly, so they will visible in several frames. However, it could be possible that a specular reflection, e.g. from the solar arrays or other shiny surfaces, generates a short light flash in the image. To exclude these events, we are analyzing whether any artificial objects cross the NELIOTA FoV.

We download orbit information about satellites in the so-called two-line element (TLE) format from two sources: (a) The database provided by the United States Strategic Command (USSTRATCOM)[2], and (b) TLE data provided by B. Gray on Github[3]. Other data sources (CelesTrak, Mike McCant) were considered but not deemed useful.

---

[2] http://www.space-track.org
[3] http://www.github.com/Bill-Gray/tles

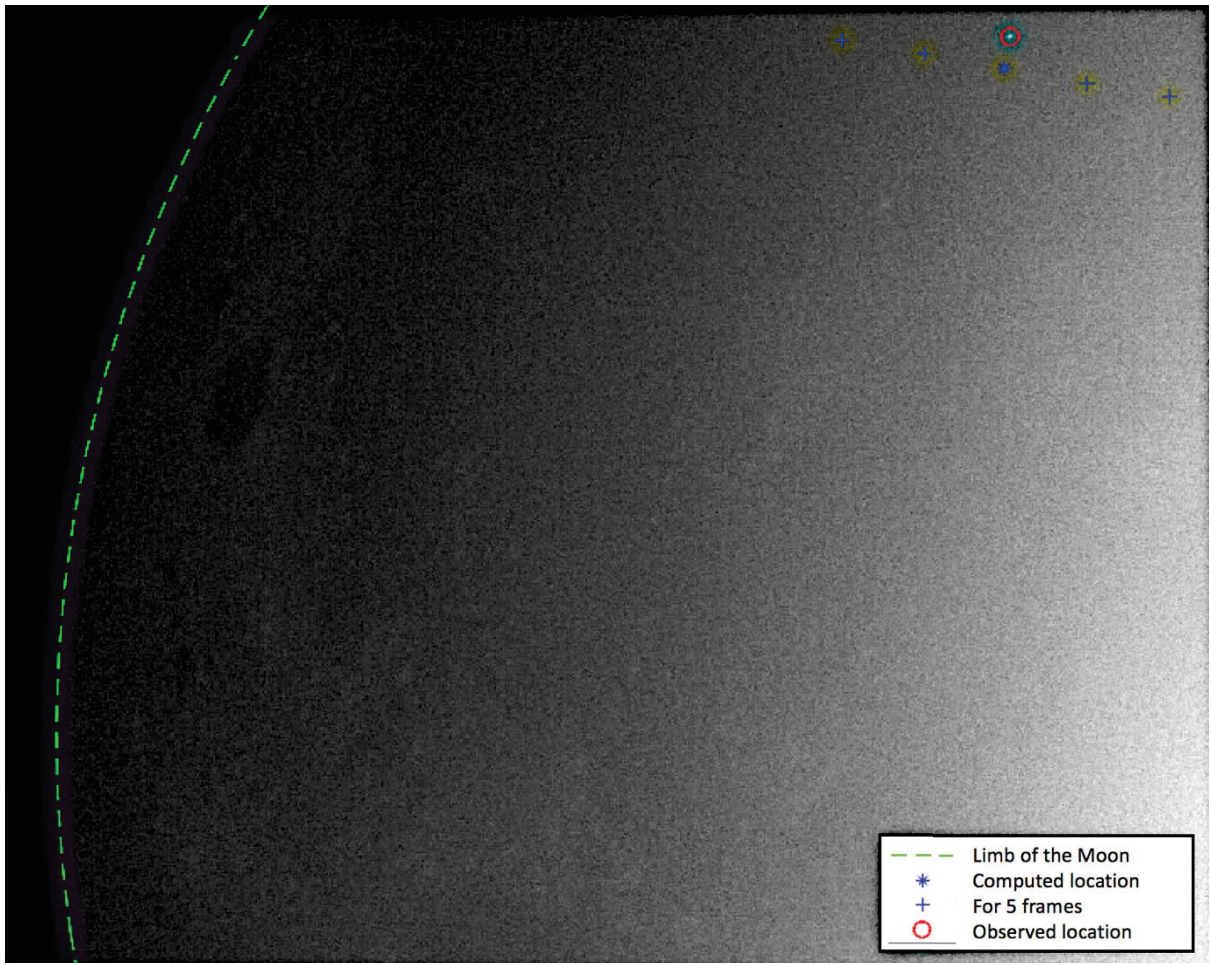

*Figure 12. Event from 01 Feb 2017, 17:27:50 UTC. Both the USSTRATCOM database and B. Gray's database give us a match for a satellite in the position shown with a blue star. The satellite is TELECOM 1B (15678).*

Data files as close as possible to the date to be checked are downloaded from the web sites. A Python script using the Simplified General Perturbations (SGP4) orbit propagator is used to propagate the elements of all available objects to the epoch of the detected event. Using JPL's SPICE[4] library, the apparent position in celestial coordinates (Right Ascension, Declination) of all objects as seen from our telescope is computed. The distance to the apparent position of the Moon is determined. If this value is smaller than a configurable threshold (set to match the size of the field of view), the satellite is flagged by the script.

The script has been tested by checking several obvious satellite detections, where an object can be seen moving through the image. As an example, the event on 01 Feb 2017, 17:27:50 UTC is presented in Fig. 12. It can be seen that the script predicts an object only several pixel away from the detection.

For all potential impact flash events, this script can be used to check whether it could possibly be a satellite. Note, however, that not finding a match does not necessarily mean that a satellite can be excluded. It may as well be that this particular satellite is simply not in the database, or does not have TLEs with a good enough accuracy. E.g., [11] checked the accuracy of propagated TLEs compared to the measured position of GPS satellites and finds that the typical deviation of the measured versus propagated in-track position is about 10 km after 5 days. For an orbit height of 800 km, this would correspond to an angular error of about 0.5°. This would already put the object outside the FoV of the instrument, thus not giving us a match. However, this check will provide more confidence about the event.

In addition, the MASTER/PROOF tool of ESA [12] was used to perform a statistical analysis. The tool was employed to find how many objects would cross the FoV of the Kryoneri telescope during a time period of 100 days, assuming a fixed pointing position. The resulting output is shown in Fig. 13. PROOF distinguishes different object types. The top-most line is the total number of crossing objects. The other lines

---

[4] https://naif.jpl.nasa.gov/naif/

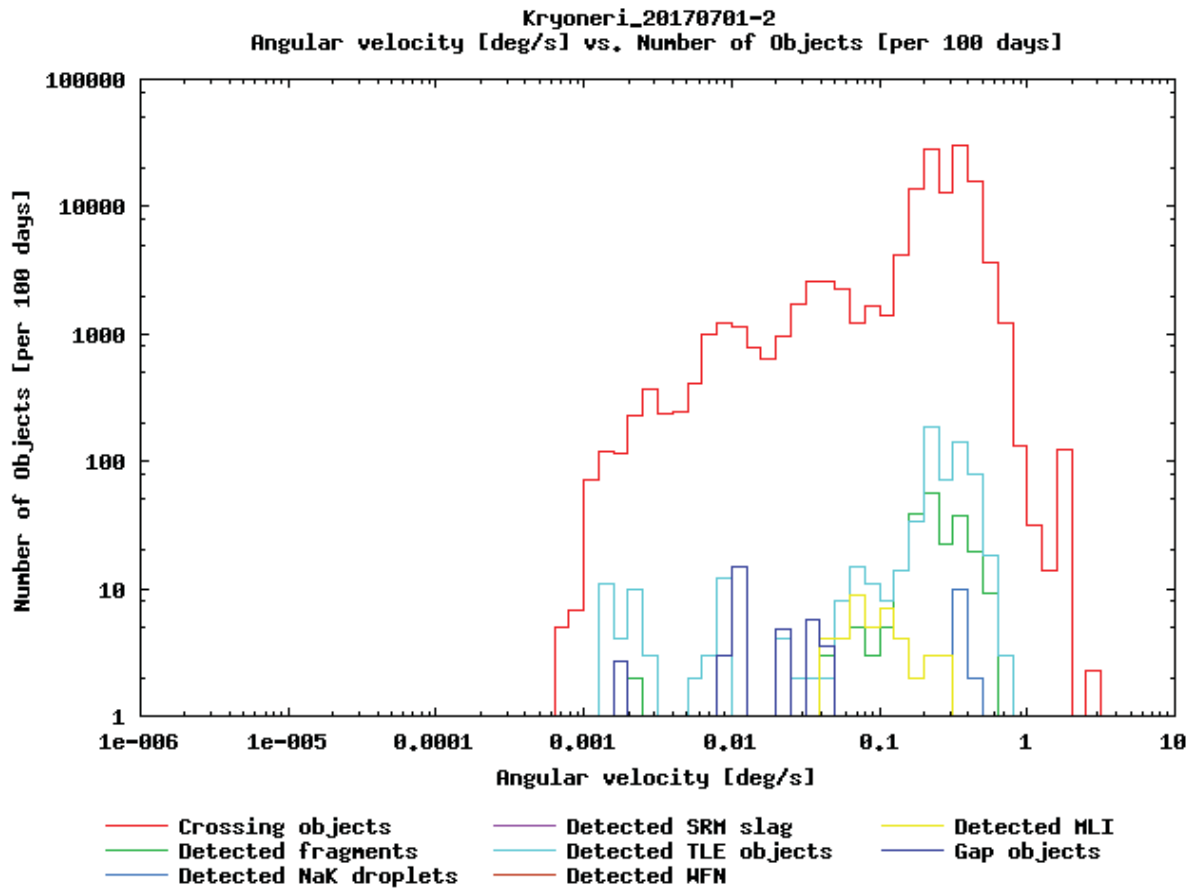

*Figure 13. Simulation of the Kryoneri telescope detections of artificial objects using ESA's MASTER/PROOF tool. 'Crossing objects' are all objects crossing the FoV in a 100-day period. Out of those only a part are bright enough to be detected.*

show the actually detected objects, taking their brightness into account.

It is assumed that a moving object can be recognized if it moves at least two pixels during the exposure time, corresponding to an angular velocity of 1.3′′/ 23 ms × 360′′/deg = 0.016°/s. Thus, only objects slower than this will not show a trail and are considered here. More details on these analyses can be found in [13].

The results so far show that during the approximately 88 hours of operations of NELIOTA the last 23 months, twenty slow moving satellites have crossed the FoV of NELIOTA during the timings of the detections of the events. Another code script, which is under the development phase, is planned to be used to check the paths of the satellites in front of the lunar disk but very near the areas where the flashes were detected. It should be noted that although for a given event a satellite match may not be found from the aforementioned data bases, it does not mean that the event is true, since these data bases only include publicly available satellite information. However, this method can be considered as a second validation of the detected flashes and will minimize the possibility of the false identification of slow moving satellites as true lunar impact flashes.

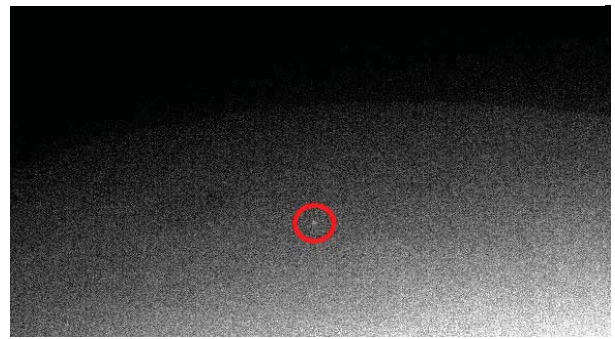

*Figure 14. The rocket body NORAD26738 (Breeze-M R/B, Russia) as detected from NELIOTA-DET on 14 Nov. 2018, 18:27:31 UTC crossing the lunar disk (inside the red circle). The object was detected in one frame of both cameras and presented round PSF similar to an impact flash.*

## 6 SUMMARY AND DISCUSSION

The present work focuses on the detection and the validation of lunar flashes observed by the NELIOTA campaign. A brief presentation of the system (hardware and software) and of the strategy followed for the observations of the lunar impact flashes is presented. Details of the software pipeline and especially for the method that is applied for the detection of possible events are also given. Furthermore, all the possible cases of detected events that can be met and the criteria used regarding their validation procedure (e.g. artefacts, true flashes) are extensively presented and are gathered for the reader's convenience in a validation flow chart (Fig. 8). In addition, existing databases for artificial satellites are checked to ensure that impact events are not confused with one of these objects.

For reasons of completeness, it is found useful to briefly present below the procedures followed for the validation of the events as well as a small sample of the current statistical results of the NELIOTA project. After the validation of the events, aperture photometry on both the true and the suspected flashes as well as on the standard stars is performed in order to derive the magnitudes of the flashes. Subsequently, using high detailed lunar maps from the software "Virtual Moon Atlas" v.6.0[5] and a special utility of the NELIOTA-DET, the selenographic coordinates of the flashes are also found. Moreover, within 24 hours of observations, the data and all the relevant information are permanently archived by the NELIOTA-ARC system in a server at NOA headquarters and are uploaded to the website[6] of the project using the NELIOTA-WEB system. Details about the photometry method can be found in [9] and [14], about the localization of the flashes in [10] and [14], and about the possible association of the projectiles with active meteoroid stream(s) at the timing of detection in [14]. Finally, statistical results including physical parameters of the projectiles, new temperatures of flashes and frequency distribution of meteoroids on the vicinity of the Earth, based on the so far results of the campaign, can be found in [14], which is planned to be submitted in the first semester of 2019. However, it was found useful to present herein a small sample of the aforementioned future work, but the present results can be considered only as preliminary. Fig. 15 shows the observations (in hrs) distribution over time for approximately the first two years of the NELIOTA campaign. It should to be noted, that the observations hours per month correspond to the "clear" hours of lunar observations i.e. the read-out time of the cameras, the time spent for the standard stars observations and the time lost due to weather and technical reasons are excluded. Fig. 16 includes the flash detections for the same time period. Finally, in Fig. 17 the locations of the so far detected flashes and the associations of the corresponding projectiles with active meteoroid streams at the time of detection are shown.

---
[5] https://www.ap-i.net/avl/en/start
[6] https://neliota.astro.noa.gr/

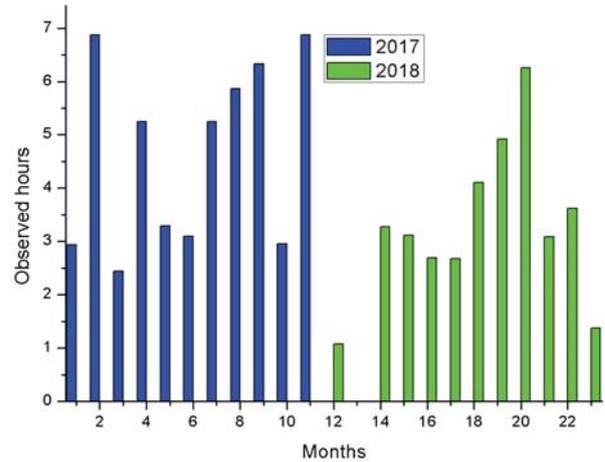

*Figure 15. Observed hours on Moon during the NELIOTA campaign.*

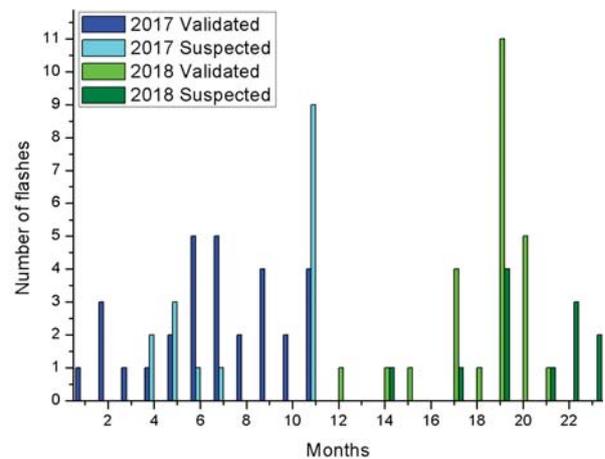

*Figure 16. Statistics of detected flashes by the NELIOTA campaign. Dark blue and light green columns denote the validated flashes, while cyan and dark green columns the suspected flashes. Note that these results are preliminary and will be further checked for possible satellite misidentification.*

To date, approximately 88 hrs of lunar observations have been obtained by the NELIOTA campaign resulted in 55 validated and another 28 suspected flashes. It should to be noted that these results are derived using the criteria described in Sect. 5 (Fig. 8). The satellite cross checking method showed that 13 validated and 7 suspected flashes from the above may have been misidentified and will be further checked. Taking into account the total observations hours, the lunar area covered by the NELIOTA setup and the number of only the validated flashes, a detection rate of $1.54 \times 10^{-7}$ (for 42 validated flashes) and $2.02 \times 10^{-7}$ flashes $hr^{-1} km^{-2}$ (for 55 validated flashes) is derived. A comparison of these values with those from previous campaigns [3], [7] shows that the NELIOTA project has almost doubled the detection rate of lunar impact flashes during its first two years of operations mainly because of the large telescope aperture used.

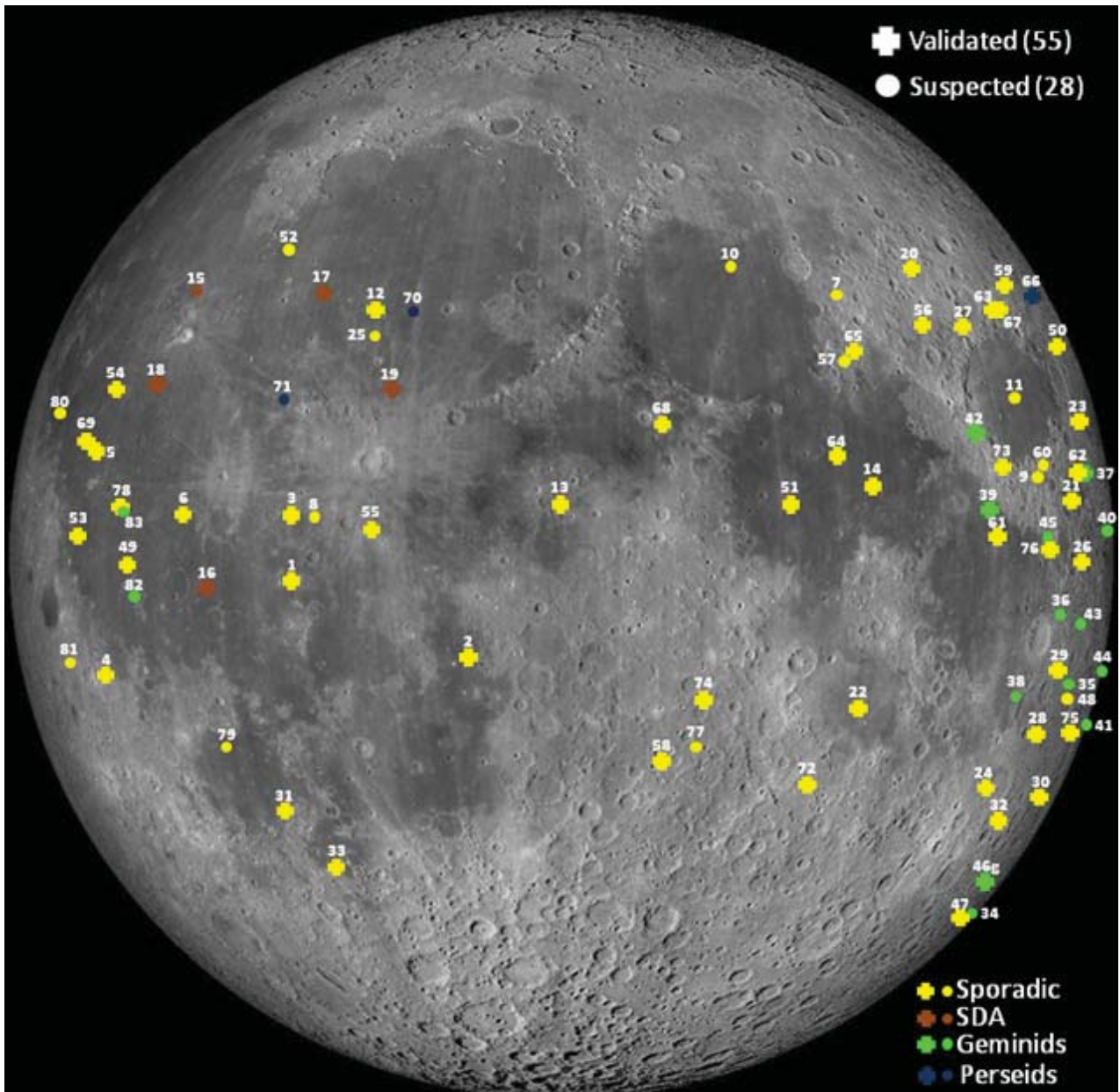

*Figure 17. Locations of the detected flashes by the NELIOTA campaign. Crosses represent the validated flashes and filled circles the suspected flashes. Different symbol colours denote the origin of the meteoroid. Yellow symbols correspond to the sporadics, while orange, green, and blue to those associated with the Southern Delta Aquarids (SDA), Geminids, and Perseids meteoroid streams, respectively (see [14] for details). The values above or by the symbols are increasing numbers and denote the chronological order of the detection. Note that these results are preliminary and will be further checked for possible satellite misidentification.*

NELIOTA operations will continue at least until January 2021 and better statistics (i.e. larger sample) about the distribution and the physical properties of the meteoroids in the inner solar system as well as about the temperatures of the impacts is anticipated to be achieved.

---